# Fractal Fluctuations in Human Walking: Comparison between Auditory and Visually Guided Stepping

PHILIPPE TERRIER[1,2]

[1]IRR, Institute for Research in Rehabilitation, Sion, Switzerland; [2]Clinique romande de réadaptation SUVACare, Sion, Switzerland

Address correspondence to Philippe Terrier, Clinique romande de réadaptation SUVACare, Av. Gd-Champsec 90, 1951 Sion Switzerland. Tel.: +41-27-603-20-77. Electronic mail: **Philippe.Terrier[at]crr-suva.ch**.

**Abstract**—In human locomotion, sensorimotor synchronization of gait consists of the coordination of stepping with rhythmic auditory cues (auditory cueing, AC). AC changes the long-range correlations among consecutive strides (fractal dynamics) into anti-correlations. Visual cueing (VC) is the alignment of step lengths with marks on the floor. The effects of VC on the fluctuation structure of walking have not been investigated. Therefore, the objective was to compare the effects of AC and VC on the fluctuation pattern of basic spatiotemporal gait parameters. Thirty-six healthy individuals walked 3 × 500 strides on an instrumented treadmill with augmented reality capabilities. The conditions were no cueing (NC), AC, and VC. AC included an isochronous metronome. In VC, projected stepping stones were synchronized with the treadmill speed. Detrended fluctuation analysis assessed the correlation structure. The coefficient of variation (CV) was also assessed. The results showed that AC and VC similarly induced a strong anti-correlated pattern in the gait parameters. The CVs were similar between the NC and AC conditions but substantially higher in the VC condition. AC and VC probably mobilize similar motor control pathways and can be used alternatively in gait rehabilitation. However, the increased gait variability induced by VC should be considered.

**Key terms**—Human locomotion, Motor control, Sensorimotor synchronization, Gait variability, Auditory cueing, Visual cueing, Long-range correlations.

**Abbreviations:**

AC:     Auditory cueing          NC:     No cueing                  VC:      Visual cueing

DFA:    Detrended fluctuation analysis          CV:      Coefficient of variation

**INTRODUCTION**

In the 1970s, Benoit Mandelbrot laid the foundations of a new method for understanding the geometry of nature. He coined the term "fractal" to describe geometric objects that look identical, regardless of the scale at which they are observed (self-similarity).[20] He also developed an analogous concept about particular time series that present self-similarities.[21] In this case, a constant statistical distribution exists across time scales. In other words, the statistical features of the parts of the series are comparable, even if the time interval during which the observations are made changes. The corollary is that these fractal time series exhibit serial correlations (or autocorrelations) between consecutive samples that slowly decrease under a power law (long-range correlations).

Biological systems are inherently complex. They are constituted of multiple sub-parts—ranging from the molecular to the population level—that interact nonlinearly on large spatial and temporal scales. Consequently, signals measured from living organisms are most often fractal. Although the classical approach considers the fluctuations in a physiological signal as random noise around the true average, the fractal analysis postulates that the structure of fluctuations enlightens us about the underlying processes that produced the signal.[10] Fractal analysis has been used on a wide variety of signals, such as heartbeat time series[29] or electroencephalograms.[1]

In human walking, the muscles of the lower limbs cyclically propel the body forward over a certain distance (step length) during a certain time (step time), thus maintaining ambulatory speed (length/time). Gait control expends low energy by delivering an optimal combination of step length and step time.[48] Furthermore, an active control of lateral foot placement is needed to provide a base of support that minimizes fall risks.[3] In short, the control of human walking is a highly complex process implying intricate interactions of feed-forward and feed-back mechanisms,[11, 35] which is a condition that is conducive for the emergence of fractal patterns in gait fluctuations. Actually, in 1995, by analyzing the walk of healthy individuals, Hausdorff and colleagues observed that the time series of stride time (i.e., the time between two consecutive heel strikes of the same limb) exhibited fractal fluctuations[16]: Deviations above and below the average tend to persist over several decades of later strides (long memory process). In order to highlight the fractal patterns, Hausdorff *et al.* used detrended fluctuation analysis (DFA), which was designed to assess the scaling properties of time series with nonstationarities.[28] In subsequent studies using DFA, they observed that, in patients with neurological gait disorders, the fractal pattern tends to be replaced by a random pattern (i.e., no correlations among successive strides).[12, 14] Thus, DFA has often been adopted to study locomotor function in patients with neurological disorders.[13]

During the decade after these seminal works, further studies using DFA extended the knowledge about fractal fluctuations in human locomotion. In 2003, West and Scaffetta[46] proposed a nonlinear stochastic dynamical model of walking that accounted for the presence of long-range correlations in stride time series. In 2005, Terrier *et al.*[42] observed that the time series of stride time, as well as the time series of stride length and stride speed were fractal. The presence of fractal patterns in several spatiotemporal gait parameters was later confirmed by Jordan *et al.* in a treadmill experiment.[18] In the meantime, theoretical considerations highlighted some issues with the DFA method,[22] questioning the presence of true long-range correlations in gait time series. However, in 2009, Delignières and Torre[5] tested the effective presence of long-range correlations in human walking with an alternative methodology[43]: by comparing different synthetic signals with actual stride time series, they concluded that artificial time series that incorporated long-range correlation by design (ARFIMA models) best fit human time series.

An interesting discovery has been that sensorimotor synchronization substantially alters the fractal structure of human walking. Sensorimotor synchronization is the coordination of movements with external rhythms or cues. This ability is responsible for unique human behavior, such as dancing or performing music.[32] In locomotion, sensorimotor synchronization implies guided stepping with external cues. The simplest expression of this synchronization behavior consists of the voluntary adjustment of heel strikes to the beat of an isochronous metronome, hereafter simply referred to as auditory cueing (AC). AC is important in gait rehabilitation. For example, in stroke patients, a recent meta-analysis of seven randomized controlled trials has concluded that the gait training with AC improves walking speed, stride length and cadence.[25] AC is also a key tool for improving gait among patients with Parkinson's disease.[45] AC has a strong effect on the fractal fluctuations of stride time but no effect on stride length and stride speed.[36, 42] The long-range correlated pattern is then replaced by anti-correlations. A value above the mean is more likely to be followed by a value below the mean (anti-persistence): the voluntary control of step duration induces a continual shift around the target (over-correction).[8, 41]

Another type of externally-driven synchronization of human movement is treadmill walking. Indeed, a motorized treadmill imposes a constant speed upon the walker. Whereas treadmill walking has only a marginal effect on the fractal fluctuation of stride time compared to overground walking,[4, 39] treadmill walking changes the serial correlations of stride speed into anti-correlations, as AC does for stride time.[8, 34, 41] Moreover, when AC is combined with treadmill walking, all the gait parameters (stride time, stride length, and stride speed) exhibit anti-correlated patterns.[34, 41] The constant speed of the treadmill (speed cueing) requires coordinated adjustments between the stride time and the stride length to maintain an appropriate speed and thus avoid falling off the treadmill. As a result, when cadence and speed are voluntarily adjusted to external cues (dual cueing), stride length must be coordinated accordingly (the "loss of redundancy" paradigm).[8, 34, 41]

In addition to the temporal adjustments of steps to auditory cues, another possible type of sensorimotor synchronization consists of adjusting step length to visual cues (hereafter referred to as visual cueing, VC). In this case, a walker anticipates the position of his or her next step to coincide with a mark on the floor. Like AC, VC has applications in gait rehabilitation. In Parkinson's disease, gait training with VC might have a long-term beneficial effect on walking capabilities.[37] Recent technical advances have led to the development of treadmills equipped with projection devices that draw visual targets on the treadmill belt (augmented reality). The use of projected visual targets on a treadmill resembles the more conventional, real, marks on the ground, but greatly facilitates the application of VC for research and rehabilitation. Promising results have been obtained with such treadmills, in particular for stroke rehabilitation.[17] In patients in the chronic stage of stroke, adaptability training using visually guided stepping improved obstacle-avoidance performance.[44] Although the field is still in its infancy, by offering complementary solutions to AC, the VC method has substantial potential for growth.

In summary, numerous studies have analyzed the effects of AC and treadmills on the variability[8, 15, 38, 39] and the fractal pattern[5, 34, 42] in human locomotion. This has led to interesting hypotheses about the neurological basis of gait control, such as the existence of a specific neural structure that generates fractal noise (the super central pattern generator hypothesis),[5, 46] or the implication of the minimum intervention principle.[8, 41] However, whether or not VC walking supports these hypotheses has yet to be investigated.

The first objective of the present study was to determine, in healthy individuals walking on a treadmill, the effects of VC on stride-to-stride fluctuations and to compare the results with the effects of AC. The hypothesis was that VC and AC similarly alter the fractal fluctuations normally present in the time series of stride time and stride length (loss of redundancy) and that the long-range correlations are assumed to be replaced by anti-correlations. The second objective was to measure the fluctuation magnitude (the coefficient of variation, CV) of the spatiotemporal gait parameters under different cueing conditions.

**MATERIALS AND METHODS**

*Subjects*

Thirty-six healthy volunteers (14 men, 22 women) with no orthopedic or neurological problems participated in the study. The mean and standard deviation (SD) of their characteristics were as follows: age 33 years (10), body height 1.72 m (0.08), and body mass 66 kg (13). The subjects had had no previous experience with walking following visual cues. All subjects signed an informed consent form according to the guidelines of the local ethic committee (Commission Cantonale Valaisanne d'Ethique Médicale, [CCVEM]), that had approved the protocol.

*Material and Experimental Procedure*

The instrumented treadmill was a C-mill model (ForceLink BV, Culemborg, The Netherlands),[17] which is equipped with an embedded vertical force platform of 70 × 300 cm. The platform recorded the vertical force and the position of the center of pressure at a sampling rate of 500 Hz. A projection system displayed visual objects on the walking area from the right side of the treadmill. *Ad hoc* control software (CueFors®) was used to compute the preliminary values of basic gait parameters (stride length and duration) and to control the projection of the visual cue drawings, the *stepping stones*."[17]

During all procedures, an elastic band was placed in front of the participant (1.40 m behind the beginning of the belt), hanging perpendicular to the handrails, at hip level. The participant was instructed to stay 10 cm behind the band, ensuring consistent placement on the walking area (approximately in the middle). The reasons were the increased safety and the standardization of the number of incoming stepping stones seen by the participant. Firstly, the preferred walking speed (PWS) of each participant was assessed using a standardized procedure,[9] which consisted of (1) a progressive increase in the treadmill speed from a low speed (2 km h$^{-1}$) until the participant reported a comfortable pace and then (2), similarly, a progressive decrease of the treadmill speed from a high speed (6 km h$^{-1}$) to a comfortable pace. The PWS was defined as the average of four tests: two with increasing speeds and two with decreasing speeds.

Then, the participant performed a 2 min walking test at PWS. The last 30 steps were analyzed to measure the average preferred stride length and stride time (i.e., the duration of one gait cycle). Then, the participant walked continuously for about 30 min at his or her PWS. Three conditions were presented in a random order. Before each condition, the experimenter described the task to the subject while he or she continued to walk. The conditions were as follows: (1) no cueing, i.e., normal walking (NC); (2) AC, i.e., walking while synchronizing the heel strike to the beep of an electronic metronome set to the preferred previously measured cadence; (3) VC, walking while adjusting steps to stepping stones, which were 20 × 30 cm moving rectangles projected on the walking area that went back at the same speed as the treadmill belt. The longitudinal distance along the belt between the successive stepping stones was set to correspond to half of the preferred stride length. The instruction was to aim accurately for each stepping stone with the foot. Given his or her position in the middle of the treadmill, the participant could see in advance two stepping stones. For a better understanding of the VC method, see the short movie provided in the supplemental material (S1). Thirty seconds of familiarization with the cueing task was given to the volunteer before the recording began. One thousand steps (500 gait cycles) were then recorded, which correspond to 10 min of walking at a step rate of 100 steps min$^{-1}$.

*Data Analysis*

One hundred and eight files (36 subjects × 3 conditions), each containing the position of the center of pressure recorded at 500 Hz, were exported for the subsequent analyses that were performed using MATLAB (MathWorks, Natick, MA). The stride time and length of each gait cycle (500 per file) were computed using a custom algorithm that implemented a validated method.[33] The reasoning behind the method is to detect heel strikes in the longitudinal signal and then to compute the distance and time between subsequent heel strikes of the same foot taking into account the treadmill speed. The average speed of each gait cycle was defined as speed = stride length/stride time. As a result, three time series (NC, AC, VC) of 500 gait cycles were obtained for stride time, stride length, and stride speed (total 3 × 3 × 36 = 324). A typical result for one participant is shown in Fig. 1.

To characterize the dispersion of the values around the mean in the time series (fluctuation magnitude), the CVs were computed, which were defined as the SD normalized by mean, and were expressed in percentages.

A fractal time series with long-range correlations exhibits an autocorrelation function C(s) that declines following a power law $C(s) \propto s^{-\gamma}, 0 < \gamma < 1$. The DFA is a method designed to assess the scaling exponent $\alpha$ ($\alpha = 1 - \gamma / 2$) in a time series with nonstationarities.[29] Although DFA may not always be appropriate to evidence the existence of long-range correlations in time series[5, 22, 43] here, I took advantage of its proved capacity to efficiently distinguish between statistical persistence (short- or long-range correlations) and anti-persistence (anti-correlations).[7, 8, 41] First, the time series of length $N$ was integrated. Then, it was divided into non-overlapping boxes of equal length $n$. In each box, a linear fit, using the least squares method, was performed. The average fluctuation $F(n)$ for that box length was:

$$F(n) = \sqrt{\frac{1}{N}\sum_{k=1}^{N}[y(k) - y_n(k)]^2}$$

where $y_n(k)$ was the y-coordinate of the k[th] point of the straight line resulting from the linear fit, and $y(k)$ was the corresponding point in the original time series. The procedure was repeated for increasing box sizes. Box sizes ($n$) ranging from 12 to 125 (i.e., N / 4) were used, with exponential spacing to avoid a bias toward larger box sizes.[23] A statistical self-affinity at different scales implies that $F(n) \approx n^{\alpha}$. Therefore, because $logF(n) \approx \alpha \cdot log(n)$, a linear fit is realized in a log-log plot between $n$ and $F(n)$ to compute the scaling exponent $\alpha$. If $\alpha$ lies between 0.5 and 1, a long-range correlation is likely. A random process (white noise) induces α value of 0.5. In the case of anti-correlation, a small $\alpha$ is expected ($\alpha < 0.5$).[40]

*Statistics*

Six dependent variables were included in the statistical analysis, namely the CV and the scaling exponent α for each gait parameter: stride time, stride length and stride speed. The independent variable was the cueing condition (NC, AC, and VC). Boxplots were used to describe the distribution of the individual results (Figs. 2 and 3). Means and SDs are shown in Tables 1 and 2. As inferential statistics, one-way, repeated-measures analyses of variance (ANOVA) were used. The explained variance was assessed with partial η2. Post hoc analyses with Tukey's honestly significant difference (HSD) tests were used to highlight specific differences between the cueing conditions; only the p-values of the significant differences have been included in Tables 1 and 2, with associated relative differences.

**RESULTS**

*Fluctuation Magnitude / Gait Variability*

The distribution of the variability results (CV) is presented in Fig. 2. The presence of some outliers can be observed. A substantially higher CV of stride time, length, and speed is observed for the VC condition. Inferential statistics (Table 1) confirm that cueing had a significant effect on the CV of the gait parameters. The ANOVA results show that cueing conditions explain a substantial part of the variance (partial $\eta^2$: 0.56–0.72). In other words, VC induced a substantial increase in the gait parameters' CV compared to NC, whereas for AC such an increase in fluctuation magnitude was not observed.

*Fractal Analysis*

The boxplots of the fractal analysis results (DFA) show that stride time and stride length exhibited long-range correlations under NC conditions ($\alpha > 0.5$), while stride speed was constantly anti-correlated ($\alpha < 0.5$). An obvious anti-correlated pattern was observed for all gait parameters in the AC and VC conditions. The ANOVA results showed significant differences among the cueing conditions for all gait parameters, with the exception of stride speed. Concerning stride length and stride time, a very large part of the variance is explained by the model (partial $\eta^2 > 0.79$), due to the change in the scaling exponent from correlation to anti-correlation. *Post hoc* analyses confirmed large changes in the correlation structure for stride length and stride time when AC and VC are compared to NC. For these parameters, no difference existed between the AC and VC conditions (similar anti-correlations).

**DISCUSSION**

Based on the analysis of a large number of gait cycles (54,000), the objective of the present study was to characterize the effects of external cues on the stride-to-stride fluctuations in gait parameters, in structural (fractal pattern) and magnitude (CV) terms. As hypothesized, concerning fractal fluctuations, spatiotemporal parameters responded similarly to AC and VC, namely, the emergence of strong anti-correlations was observed. However, AC and VC did not affect the fluctuation magnitude in the same way: VC induced a substantial increase in the gait parameters' CVs compared to the control condition, whereas no differences were observed for AC.

When individuals walk in a constant environment, basic spatiotemporal parameters fluctuate from stride-to-stride in a narrow window, very likely framed by energetic, biomechanical,[26, 48] and stability[27] constraints. In overground walking, typical CV values between 2.5% and 3% have been reported in healthy young subjects.[39, 42] Similar results have been observed in treadmill walking,[39, 41] which were confirmed in the present study (CV 1.8% to 2.8%, Table 1). The effect of AC on the fluctuation magnitude has been assessed in experimental and clinical studies. In elderly subjects, in two studies, which analyzed overground walking with or without AC given at a preferred cadence, a small increase in gait variability[15] or no effect[47] was found. In a previous study that combined treadmill and AC ($N$ = 20 young adults, 5 min walking), when the AC and NC conditions were compared, a small decrease in the CVs for the stride time (−19%), stride length (−17%), and stride speed (−7%) has been observed.[41] Compared to that study, the present study included more participants ($N$ = 36) and longer trials (10 min). The (not significant) changes induced by AC were stride time, −13%; stride length, −7%; and stride speed, 0% (Table 1). In summary, AC has little impact on stride-to-stride fluctuation magnitude. As illustrated in Fig.1, the erratic wandering of the parameter around the mean, which is typical of fractal fluctuation (first column), is replaced by high frequency noise around the imposed frequency, which is typical of anti-correlated patterns (second column). Thus, these concomitant changes in the fluctuation pattern triggered by AC, which decreased low frequency wanderings and increased high frequency oscillations, nearly compensate for each other to give a comparable fluctuation magnitude. On the other hand, VC had a profound effect on the fluctuation magnitude. The stride time CV increased by 51%, stride length CV by 73%, and stride speed CV by 63% (Table 1). Although the participants were able to target the stepping stones, their stride length oscillated in an extended range (CV: 5%, 7 cm for a 1.3 m stride length). Due to the interdependence among gait parameters, the stride time and the stride speed were also affected.  This suggests that VC was a challenging task. A cause of the difficulty was perhaps that only two incoming steps were seen in advance by the subject, which could be too short a warning to anticipate precise foot placement. Peper et al.[30], using the same treadmill as in the present study, showed that individuals spontaneously chose a lower walking speed in VC condition: this corroborates the hypothesis that a sufficient reaction time to the incoming targets should be allowed to help the subject to be comfortable with VC. However, further studies are needed to analyze gait variability at different speeds under VC condition. As far as I know, this strong effect on fluctuation magnitude when VC is combined with treadmill walking has not been described in the literature and has to be confirmed independently. Interestingly, high variability has been observed in the pathological gait of patients with Parkinson's disease (CV of stride length: 5.32%).[19] Patients with Parkinson's disease rely more on vision while walking.[2] The fact that healthy individuals, who guide their steps based on vision, also exhibited high variability may have important implications in fundamental research and clinical application, which deserve further investigations.

Walking on a treadmill requires coordinated control of stride time and stride length to match the treadmill speed. As evidence of this coordinated regulation, several studies have shown that stride time and stride length are actually cross-correlated:[7, 34, 41] In other words, stride time and stride length vary over time in a similar way (positive correlations).[34] A model that explains how motor control manages speed regulation has been proposed by Dingwell and collaborators.[6, 8] A key feature of this model is that an infinite combination of stride time and stride length is possible to meet the goal of maintaining a constant speed. In other words, redundancy exists between spatial and temporal control of walking speed: if a deviation occurs in the stride length, the deviation can be compensated for by a correction in the stride time, and vice versa. As a result, deviations can persist over consecutive strides, which may explain long-range correlations and fractal fluctuations (the minimum intervention principle).[6] If two goals are simultaneously imposed on a walking individual, for instance, treadmill walking (speed goal) and AC (stride time goal), the redundancy disappears, and tight regulation of stride length is also performed, which leads to the loss of the fractal fluctuations, as demonstrated in previous studies[8, 34, 41] and in the present study (Fig. 3). The main goal of the present study was to provide further evidence to support this model. As expected, imposing a spatial goal (stepping to visual cues) modified the fluctuation pattern of stride length as well as stride time (Fig. 3 and Table 2). Actually, AC and VC altered fractal fluctuations in a similar way (Table 2). As observed previously, strong anti-correlations appeared.[41, 42]

As illustrated in Table 3, the results of the present study provide new information to complement what is already known about voluntary synchronization of gait to external cues. The main gap that remains to be filled is the fluctuation pattern of overground walking under the VC condition. It can be assumed that stride length would then be anti-correlated, while stride time and stride speed would remain correlated. However, conducting such an experiment is technically challenging.

Taken together, these findings highlight a stereotyped response of gait control to external cueing, the neurobiological basis of which remains to be elucidated. What is already known is that sensorimotor synchronization requires anticipation to align motor response with external cues. The delivery of sensory signals and central processing delay the motor response. Thus, to get in time, motor command must anticipate future steps based on internal models and past sensory inputs and not react to current stimuli.[24] This anticipation leads to a well-known phenomenon in rhythmic tapping experiments: the taps precede the metronome (negative mean asynchrony).[31] Likewise, in gait experiments with AC, the heel-strike slightly precedes (50 ms on average) the next occurrence of the metronome sound.[23] The anticipation of movement based on visual cues, for example, reaching a moving object, is also a basic task of sensorimotor coordination that implies feedforward mechanisms.[24] In the present experiment, the subject saw two stepping stones in advance to anticipate the correct step lengths. In short, guided stepping requires voluntary adaptation to external cues and anticipation to produce a timed motor response. It can be assumed that visual, auditory, and somatosensory afferents converge to modulate a central pacemaker that triggers the appropriate rhythmic gait behavior under voluntary control. How exactly this sensory feedback feeds this hypothetical pacemaker, and how an anti-correlated pattern is produced remain to be further investigated.

In conclusion, because it can be assumed that AC and VC mobilize the same motor pathways, they can probably be used alternatively in gait rehabilitation. The efficiency of VC to enhance walking abilities in patients with neurological gait disorders needs further studies. However, the high gait variability induced by VC might have detrimental effects, for instance, a lower dynamic balance. This should be taken into account in the development of VC rehabilitation methods.

**ACKNOWLEDGMENTS**

The author warmly thanks Emilie Du Fay de Lavallaz for his valuable support in bibliographical research and Vincent Bonvin for his help in data collection. The study was funded by the Swiss accident insurance company SUVA, an independent, non-profit company under public law, and by the clinique romande de réadaptation. The IRR (Institute for Research in Rehabilitation) is funded by the State of Valais and the City of Sion. Study funders had no role in the collection, analysis, and interpretation of data; in the writing of the manuscript; and in the decision to submit the manuscript for publication.

# REFERENCES

1.	Accardo A., M. Affinito, M. Carrozzi and F. Bouquet. Use of the fractal dimension for the analysis of electroencephalographic time series. *Biol. Cybern.* 77:339-350, 1997.

2.	Azulay J.-P., S. Mesure, B. Amblard, O. Blin, I. Sangla and J. Pouget. Visual control of locomotion in Parkinson's disease. *Brain* 122:111-120, 1999.

3.	Bauby C. E. and A. D. Kuo. Active control of lateral balance in human walking. *J. Biomech.* 33:1433-1440, 2000.

4.	Chang M. D., S. Shaikh and T. Chau. Effect of treadmill walking on the stride interval dynamics of human gait. *Gait Posture* 30:431-435, 2009.

5.	Delignieres D. and K. Torre. Fractal dynamics of human gait: a reassessment of the 1996 data of Hausdorff et al. *J. Appl. Physiol.* 106:1272-1279, 2009.

6.	Dingwell J. B. and J. P. Cusumano. Identifying Stride-To-Stride Control Strategies in Human Treadmill Walking. *PloS ONE* 10:e0124879, 2015.

7.	Dingwell J. B. and J. P. Cusumano. Re-interpreting detrended fluctuation analyses of stride-to-stride variability in human walking. *Gait Posture* 32:348-353, 2010.

8.	Dingwell J. B., J. John and J. P. Cusumano. Do humans optimally exploit redundancy to control step variability in walking? *PLoS Comput. Biol.* 6:e1000856, 2010.

9.	Dingwell J. B. and L. C. Marin. Kinematic variability and local dynamic stability of upper body motions when walking at different speeds. *J. Biomech.* 39:444-452, 2006.

10.	Eke A., P. Herman, L. Kocsis and L. Kozak. Fractal characterization of complexity in temporal physiological signals. *Physiol. Meas.* 23:R1, 2002.

11.	Hamacher D., F. Herold, P. Wiegel, D. Hamacher and L. Schega. Brain activity during walking: a systematic review. *Neurosci. Biobehav. R.* 57:310-327, 2015.

12.	Hausdorff J. M. Gait dynamics in Parkinson's disease: common and distinct behavior among stride length, gait variability, and fractal-like scaling. *Chaos* 19:026113, 2009.

13.	Hausdorff J. M. Gait dynamics, fractals and falls: finding meaning in the stride-to-stride fluctuations of human walking. *Hum. Mov. Sci.* 26:555-589, 2007.

14.	Hausdorff J. M., A. Lertratanakul, M. E. Cudkowicz, A. L. Peterson, D. Kaliton and A. L. Goldberger. Dynamic markers of altered gait rhythm in amyotrophic lateral sclerosis. *J. Appl. Physiol.* 88:2045-2053, 2000.

15.	Hausdorff J. M., J. Lowenthal, T. Herman, L. Gruendlinger, C. Peretz and N. Giladi. Rhythmic auditory stimulation modulates gait variability in Parkinson's disease. *Eur. J. Neurosci.* 26:2369-2375, 2007.

16.	Hausdorff J. M., C. K. Peng, Z. Ladin, J. Y. Wei and A. L. Goldberger. Is walking a random walk? Evidence for long-range correlations in stride interval of human gait. *J. Appl. Physiol.* 78:349-358, 1995.

17.	Heeren A., M. W. van Ooijen, A. C. Geurts, B. L. Day, T. W. Janssen, P. J. Beek, M. Roerdink and V. Weerdesteyn. Step by step: a proof of concept study of C-Mill gait adaptability training in the chronic phase after stroke. *J. Rehabil. Med.* 45:616-622, 2013.

18.	Jordan K., J. H. Challis and K. M. Newell. Walking speed influences on gait cycle variability. *Gait Posture* 26:128-134, 2007.

19.     Lewis G. N., W. D. Byblow and S. E. Walt. Stride length regulation in Parkinson's disease: the use of extrinsic, visual cues. *Brain* 123 ( Pt 10):2077-2090, 2000.

20.     Mandelbrot B. B. *Les objets fractals: forme, hasard et dimension*. Flammarion, 1975.

21.     Mandelbrot B. B. and J. W. Van Ness. Fractional Brownian motions, fractional noises and applications. *SIAM review* 10:422-437, 1968.

22.     Maraun D., H. W. Rust and J. Timmer. Tempting long-memory - on the interpretation of DFA results. *Nonlinear Proc. Geoph.* 11:495-503, 2004.

23.     Marmelat V., K. Torre, P. J. Beek and A. Daffertshofer. Persistent fluctuations in stride intervals under fractal auditory stimulation. *PLoS ONE* 9:e91949, 2014.

24.     Miall R. C. and D. M. Wolpert. Forward models for physiological motor control. *Neural networks* 9:1265-1279, 1996.

25.     Nascimento L. R., C. Q. de Oliveira, L. Ada, S. M. Michaelsen and L. F. Teixeira-Salmela. Walking training with cueing of cadence improves walking speed and stride length after stroke more than walking training alone: a systematic review. *J. Physiother.* 61:10-15, 2015.

26.     O'Connor S. M., H. Z. Xu and A. D. Kuo. Energetic cost of walking with increased step variability. *Gait Posture* 36:102-107, 2012.

27.     Owings T. M. and M. D. Grabiner. Variability of step kinematics in young and older adults. *Gait Posture* 20:26-29, 2004.

28.     Peng C.-K., S. V. Buldyrev, S. Havlin, M. Simons, H. E. Stanley and A. L. Goldberger. Mosaic organization of DNA nucleotides. *Phys. Rev. E* 49:1685, 1994.

29.     Peng C. K., S. Havlin, H. E. Stanley and A. L. Goldberger. Quantification of scaling exponents and crossover phenomena in nonstationary heartbeat time series. *Chaos* 5:82-87, 1995.

30.     Peper C. L. E., M. J. de Dreu and M. Roerdink. Attuning one's steps to visual targets reduces comfortable walking speed in both young and older adults. *Gait Posture* 41:830-834, 2015.

31.     Repp B. H. Sensorimotor synchronization: a review of the tapping literature. *Psychon. Bull. Rev.* 12:969-992, 2005.

32.     Repp B. H. and Y.-H. Su. Sensorimotor synchronization: A review of recent research (2006-2012). *Psychon. Bull. Rev.* 20:403-452, 2013.

33.     Roerdink M., B. H. Coolen, B. H. Clairbois, C. J. Lamoth and P. J. Beek. Online gait event detection using a large force platform embedded in a treadmill. *J. Biomech.* 41:2628-2632, 2008.

34.     Roerdink M., A. Daffertshofer, V. Marmelat and P. J. Beek. How to Sync to the Beat of a Persistent Fractal Metronome without Falling Off the Treadmill? *PLoS ONE* 10:e0134148, 2015.

35.     Rossignol S., R. Dubuc and J. P. Gossard. Dynamic sensorimotor interactions in locomotion. *Physiol. Rev.* 86:89-154, 2006.

36.     Sejdic E., Y. Fu, A. Pak, J. A. Fairley and T. Chau. The Effects of Rhythmic Sensory Cues on the Temporal Dynamics of Human Gait. *PLoS ONE* 7:e43104, 2012.

37.     Sidaway B., J. Anderson, G. Danielson, L. Martin and G. Smith. Effects of long-term gait training using visual cues in an individual with Parkinson disease. *Phys. Ther.* 86:186-194, 2006.

38.     Terrier P. Step-to-Step Variability in Treadmill Walking: Influence of Rhythmic Auditory Cueing. *PLoS ONE* 7:e47171, 2012.

39.     Terrier P. and O. Dériaz. Kinematic variability, fractal dynamics and local dynamic stability of treadmill walking. *J. Neuroeng. Rehabil.* 8:12, 2011.

40.     Terrier P. and O. Dériaz. Nonlinear dynamics of human locomotion: effects of rhythmic auditory cueing on local dynamic stability. *Front. Physiol.* 4:230, 2013.

41.     Terrier P. and O. Dériaz. Persistent and anti-persistent pattern in stride-to-stride variability of treadmill walking: influence of rythmic auditory cueing. *Hum.Mov. Sci.* 31:1585-1597, 2012.

42.     Terrier P., V. Turner and Y. Schutz. GPS analysis of human locomotion: further evidence for long-range correlations in stride-to-stride fluctuations of gait parameters. *Hum. Mov. Sci.* 24:97-115, 2005.

43.     Torre K., D. Delignieres and L. Lemoine. Detection of long-range dependence and estimation of fractal exponents through ARFIMA modelling. *Brit. J. Math. Stat. Psy.* 60:85-106, 2007.

44.     van Ooijen M. W., A. Heeren, K. Smulders, A. C. Geurts, T. W. Janssen, P. J. Beek, V. Weerdesteyn and M. Roerdink. Improved gait adjustments after gait adaptability training are associated with reduced attentional demands in persons with stroke. *Exp. Brain Res.* 233:1007-1018, 2015.

45.     van Wegen E. E., M. A. Hirsch, M. Huiskamp and G. Kwakkel. Harnessing Cueing Training for Neuroplasticity in Parkinson Disease. *Top. Geriatr. Rehabil.* 30:46-57, 2014.

46.     West B. J. and N. Scafetta. Nonlinear dynamical model of human gait. *Phys. Rev. E* 67:051917, 2003.

47.     Wittwer J. E., K. E. Webster and K. Hill. Music and metronome cues produce different effects on gait spatiotemporal measures but not gait variability in healthy older adults. *Gait Posture* 37:219-222, 2013.

48.     Zarrugh M. Y., F. N. Todd and H. J. Ralston. Optimization of energy expenditure during level walking. *Eur. J. Appl. Physiol. Occup. Physiol.* 33:293-306, 1974.

# Figures

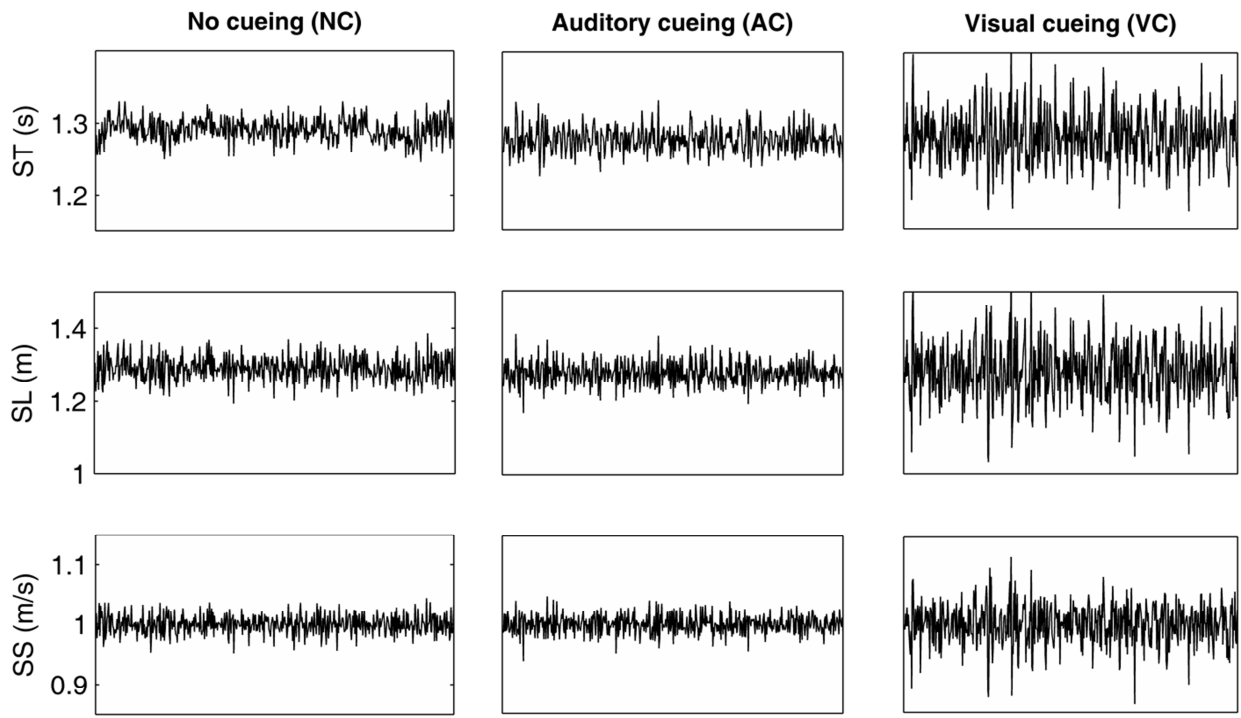

**FIGURE 1.** Typical results. One participant walked approximately 3X10min on an instrumented treadmill under 3 conditions: no cueing (NC), auditory cueing (AC) and visual cueing (VC). In each condition, 500 strides were recorded (x-axes, #stride). Basic spatiotemporal gait parameters (y-axes), i.e., stride time (ST), stride length (SL), and stride speed (SS), were assessed.

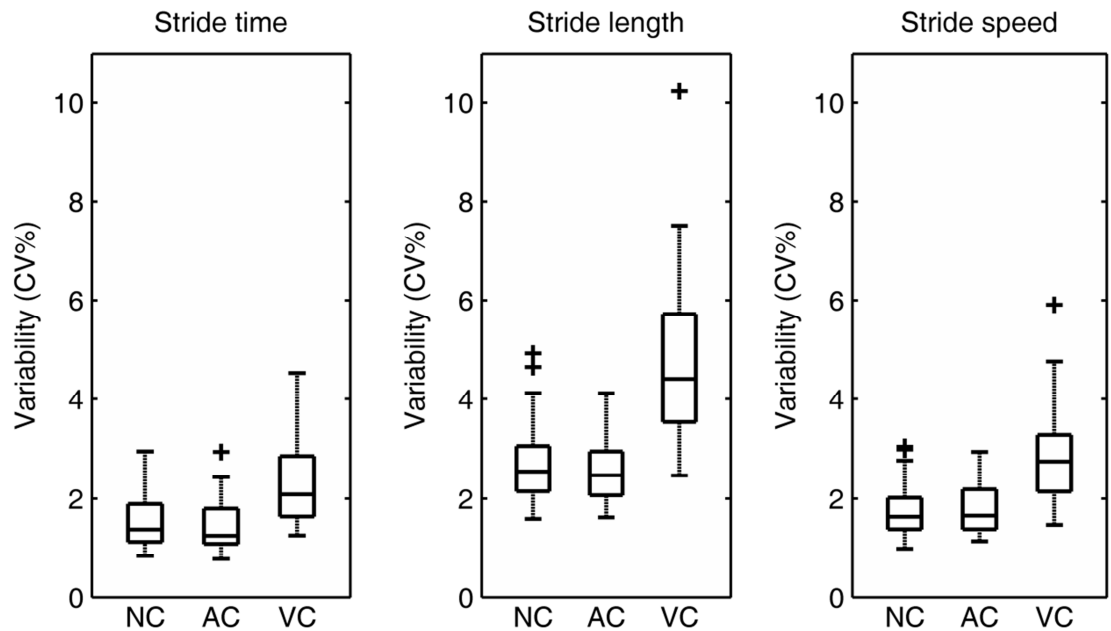

**FIGURE 2.** Boxplots of the fluctuation magnitude results. Thirty-six subjects walked on a treadmill under 3 conditions: NC = no cueing (normal walking); AC = auditory cueing (metronome walking); VC = visual cueing (visually guided stepping). In each condition for each subject, 500 gait cycles were recorded, from which the relative fluctuation magnitude (variability) of the spatiotemporal gait parameters was assessed (i.e. CV = SD / mean * 100). Boxplots show the quartiles, the medians and the ranges of the individual results. Outliers are indicated with + signs.

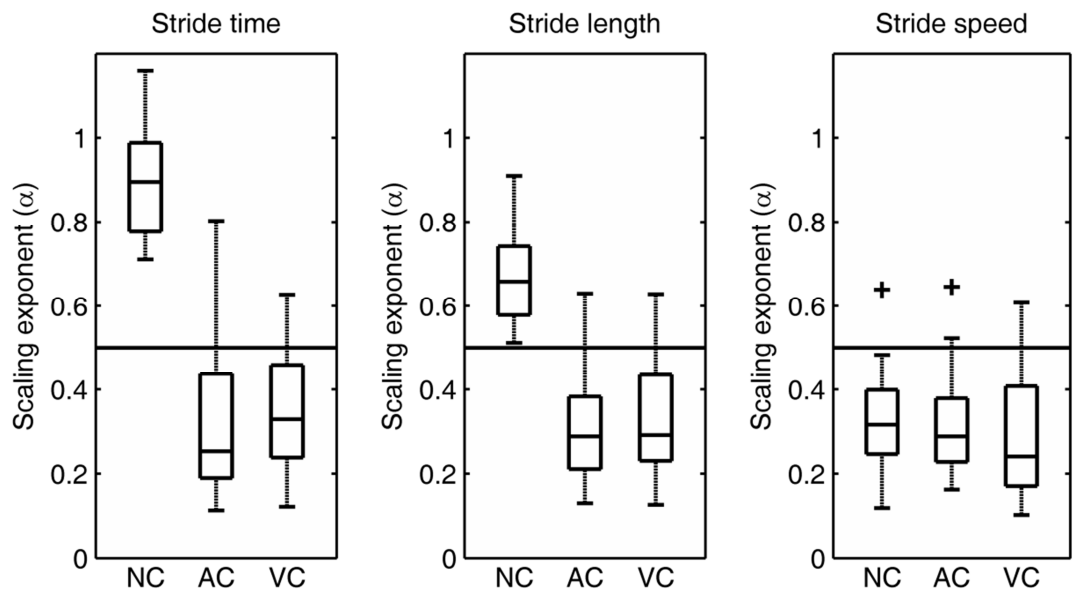

**FIGURE 3.** Boxplots of the fractal fluctuation results. Thirty-six subjects walked on a treadmill under 3 conditions: NC = no cueing (normal walking); AC = auditory cueing (metronome walking); VC = visual cueing (visually guided stepping). In each condition for each subject, 500 gait cycles were recorded, from which the serial correlations of the spatiotemporal gait parameters were assessed (scaling exponent α, DFA). Boxplots show the quartiles, the medians and the ranges of the individual results. Outliers are indicated with + signs.

# Tables

**TABLE 1. Fluctuation magnitude of the gait parameters**

| N=36 | Descriptive statistics: Mean (SD) | | | | | | ANOVA | | | | Post hoc comparisons | | | |
|---|---|---|---|---|---|---|---|---|---|---|---|---|---|---|
| | NC | | AC | | VC | | F | p | $\eta^2$ | CI | VC vs NC | | VC vs AC | |
| | | | | | | | | | | | p | Δ% | p | Δ% |
| Stride time CV (%) | 1.6 | (0.6) | 1.4 | (0.5) | **2.3** | (0.9) | 45.1 | **<0.001** | 0.56 | 0.45 | 0.70 | **<0.001** | 51 | **<0.001** | 66 |
| Stride length CV (%) | 2.8 | (0.8) | 2.6 | (0.7) | **4.8** | (1.7) | 88.5 | **<0.001** | 0.72 | 0.62 | 0.83 | **<0.001** | 73 | **<0.001** | 82 |
| Stride speed CV (%) | 1.8 | (0.5) | 1.8 | (0.5) | **2.8** | (0.9) | 83.6 | **<0.001** | 0.70 | 0.60 | 0.83 | **<0.001** | 63 | **<0.001** | 62 |

Thirty-six subjects walked on a treadmill under 3 conditions: NC = no cueing (normal walking); AC = auditory cueing (metronome walking); and VC = visual cueing (visually guided stepping). In each condition for each subject, 500 gait cycles were recorded, from which the relative fluctuation magnitude (variability) of the spatiotemporal gait parameters was assessed (i.e. CV = SD/mean * 100). Descriptive and inferential statistics and significant post hoc comparisons are shown. Δ% is the relative difference: (VC-NC)/NC *100. Significant results are shown in bold.

**TABLE 2. Long-range correlations (fractal fluctuations) of the gait parameters**

| N=36 | Descriptive statistics: Mean (SD) | | | | | | ANOVA | | | | Post hoc comparisons | | | |
| --- | --- | --- | --- | --- | --- | --- | --- | --- | --- | --- | --- | --- | --- | --- |
| | NC | | AC | | VC | | F | p | $\eta^2$ | $\eta^2$CI | AC vs NC | | VC vs NC | |
| | | | | | | | | | | | p | $\Delta$% | p | $\Delta$% |
| Stride time $\alpha$ | 0.90 | (0.13) | 0.32 | (0.17) | 0.34 | (0.14) | 209 | **<0.001** | 0.86 | 0.80 0.91 | **<0.001** | -64 | **<0.001** | -62 |
| Stride length $\alpha$ | 0.67 | (0.12) | 0.31 | (0.12) | 0.33 | (0.15) | 135 | **<0.001** | 0.79 | 0.73 0.86 | **<0.001** | -54 | **<0.001** | -51 |
| Stride speed $\alpha$ | 0.32 | (0.11) | 0.32 | (0.11) | 0.29 | (0.15) | 1.43 | 0.25 | 0.04 | 0.00 0.21 | - | - | - | - |

Thirty-six subjects walked on a treadmill under 3 conditions: NC = no cueing (normal walking); AC = auditory cueing (metronome walking); VC = visual cueing (visually guided stepping). In each condition for each subject, 500 gait cycles were recorded, from which the serial correlations of the spatiotemporal gait parameters were assessed (scaling exponent α, detrended fluctuation analysis). Descriptive and inferential statistics and significant post hoc comparisons are shown. ◌% is the relative difference: (VC-NC)/NC *100. Significant results are shown in bold.

**TABLE 3. Overview of the cueing effects on fractal dynamics and fluctuations magnitude in human walking.**

| | Stride time | Stride length | Stride speed |
|---|---|---|---|
| Overground walking[16, 42] | α >0.5 / ↓ | α >0.5 / ↓ | α >0.5 / ↓ |
| Overground walking + AC[42] | α <0.5 / ↓ | α >0.5 / ↓ | α >0.5 / ↓ |
| Overground walking + VC | ? | ? | ? |
| Treadmill walking[8, 34, 41] | α >0.5 / ↓ | α >0.5 / ↓ | α <0.5 / ↓ |
| Treadmill walking + AC[34, 41] | α <0.5 / ↓ | α <0.5 / ↓ | α <0.5 / ↓ |
| Treadmill walking + VC | **α <0.5 / ↑** | **α <0.5 / ↑** | **α <0.5 / ↑** |

Each row shows the results for a particular cueing condition on the different gait parameters (columns). AC: auditory cueing. VC: visual cueing. α >0.5 (green): DFA indicates the presence of long-range correlations. α <0.5 (red): DFA indicates the presence of anti-correlations. ↓ (green): low fluctuation magnitude. ↑ (red): high fluctuation magnitude. The new findings brought by the present study are shown in bold. The numbers refers to the bibliography at the end of the article.